\documentclass[sigconf, 10pt, nonacm]{acmart}
\title{Bandwidth, Latency, and 400 Million Kilometers: The Case for Mars-Local Compute}

\author{Maleeha Masood}
\affiliation{
  \institution{University of Illinois Urbana-Champaign}
  \country{Urbana, IL, USA}
}
\author{Indranil Gupta}
\affiliation{
  \institution{University of Illinois Urbana-Champaign}
  \country{Urbana, IL, USA}
}
\author{Deepak Vasisht}
\affiliation{
  \institution{University of Illinois Urbana-Champaign}
  \country{Urbana, IL, USA}
}

\newcommand{\para}[1]{\vspace{6pt}\textbf{#1}}
\usepackage{booktabs}
\usepackage{tabularx}
\usepackage{array}

\newcolumntype{Y}{>{\raggedright\arraybackslash}X}

\begin{abstract}

There have been recent proposals for human settlements on Mars in 2030s~\cite{spacex_mars_2026,dinkel2024vela,vora2026mars}. Any human activity on Mars must be preceded by extensive robotic exploration. However, Mars exploration is bottlenecked by the low bandwidth, intermittent Mars--Earth link. For example, HiRISE, a high-resolution camera onboard the Martian orbiter MRO imaged less than 3\% of Mars over eleven years, even though MRO's low resolution Context Camera had mapped more than 99\% of Mars in that time. We present a systems case for shared compute for Mars exploration. Such Mars-local compute, paired with advances in computer vision and AI, can enable large volumes of data to be collected and processed on Mars while sending periodic updates, insights, and selective datasets to Earth. To overcome the lack of surface infrastructure on Mars,  we propose a two-tier in-orbit deployment of computational satellites that provides consistent coverage and bandwidth. Our analysis shows that the proposed deployment can start small: one areostationary node makes compute reachable from all active Mars missions, two additional areostationary nodes can extend this coverage to roughly 90\% of the planet, while low-Mars-orbit nodes add high-rate surface links and compute capacity where demand grows.

\end{abstract}

\begin{document}

\maketitle

\section{Introduction}

Human exploration of Mars is moving from long-term aspiration towards operational preparation. In July 2026, NASA began recruiting participants for a one-year Mars mission simulation
\cite{qadri2026simulation,nasa_hrp_moon_mars_simulation}. This 
aspiration builds on decades of advances in robotic reconnaissance. 
Orbiters have mapped candidate landing sites and monitored weather and dust activity, while rovers have characterized local terrain and surface conditions.

This reconnaissance campaign is constrained less by what its instruments can observe than by how much of the resulting data can be returned to Earth. HiRISE, a camera aboard the Mars Reconnaissance Orbiter (MRO), can acquire as much as 28 gigabits in six seconds \cite{hirise}. Yet, MRO returns data to Earth at rates of only a few megabits per second. 
Even at its maximum reported downlink rate of 6 Mbps, transmitting a single 28 gigabit acquisition would take more than an hour. High-resolution imaging is therefore necessarily selective: NASA reported in 2017 that HiRISE had imaged only about 3\% of Mars, even though MRO’s lower-resolution Context Camera had mapped more than 99\% of the planet since 2006 \cite{mapmars}.

High communication latency also limits how quickly Earth-based operators can respond to changing conditions on Mars. Depending on the relative positions of Earth and Mars, the planets can be 400 million kilometers apart and one-way communication delays range from approximately 3 to 22 minutes. Communication can also be blocked for as long as 2 weeks when the Sun lies between Earth and Mars, and disrupts radio communication~\cite{nasa2024marsarchitecture,nasa_adaptive_systems}. 
More importantly, routine rover operations are organized around daily planning cycles: observations are transmitted to Earth for analysis, and a new activity plan is typically uploaded once per sol (i.e. one day on Mars). As a result, decisions that require Earth-based interpretation may not be executed until the following sol \cite{marsplanning}.

Mars therefore represents an extreme form of a familiar distributed-systems problem characterized by three features: (i) instruments generate data faster than the backhaul can carry it, (ii) connectivity is delayed and intermittent, and (iii) centralized decisions arrive too slowly to guide collection interactively. On Earth, such constraints are solved by pushing compute closer to the source of data collection either through placing more compute on edge devices or through nearby datacenters. Placing more compute on edge devices increases weight, size, and power requirements. Furthermore, surface datacenters cannot be shared across devices due to the lack of networking and communication infrastructure on Mars.

We propose a two-tier computing infrastructure in Mars orbit. One to three areostationary\footnote{analogous to Earth's geostationary satellites} nodes provide persistent access, shared state, and coordination across 90\% of the planet. These areostationary nodes are paired with low-Mars-orbit nodes to create high-rate surface links and additional compute where demand grows. Together, these tiers form a small, incrementally deployable computing infrastructure in Mars orbit. The proposed infrastructure could open up a whole host of exploration forms impossible today. For instance, 
missions could fuse observations across rovers and orbiters, maintain shared maps and models, and trigger follow-up measurements before waiting for the next Earth planning cycle. As Mars is predicted to host increasingly more sensors, vehicles, aircraft, and eventually human crews, these assets could share persistent computing services rather than operate as isolated missions. Finally, when humans arrive on Mars, this network can be re-purposed to serve both human and robotic explorers.

This work is a starting point for computing at Mars, not a complete blueprint. We provide a first-order architecture, feasibility analysis, and estimates for the community to discuss, refine, or challenge. We hope this paper jump starts  a broader conversation about building compute on Mars, the new forms of exploration it could enable, and the challenges that must be solved to make it practical.

\section{Primer and Related Work}

\subsection{The Current Martian Fleet}

\begin{table}
\centering
\footnotesize
\setlength{\tabcolsep}{3pt}
\renewcommand{\arraystretch}{0.95}
\begin{tabularx}{\columnwidth}{@{}lYY@{}}
\toprule
 & \textbf{Terrestrial LEO} & \textbf{Mars Orbit} \\
\midrule
Earth control loop
    & milliseconds
    & 3--22 min one way \\
Ground infrastructure
    & dense gateways and cloud
    & no surface backbone \\
Earth--Link Outage
    & brief or reroutable
    & periodic and unavoidable \\
Solar irradiance
    & 1.36 kW/m$^2$
    & 0.59 kW/m$^2$ \\
Launch windows
    & frequent
    & every 26 months \\
\bottomrule
\end{tabularx}
\caption{\centering Mars operates under different assumptions from terrestrial LEO. 
}
\label{tab:leo-vs-mars}
\end{table}
\begin{figure}
    \centering
    \includegraphics[width=\linewidth]{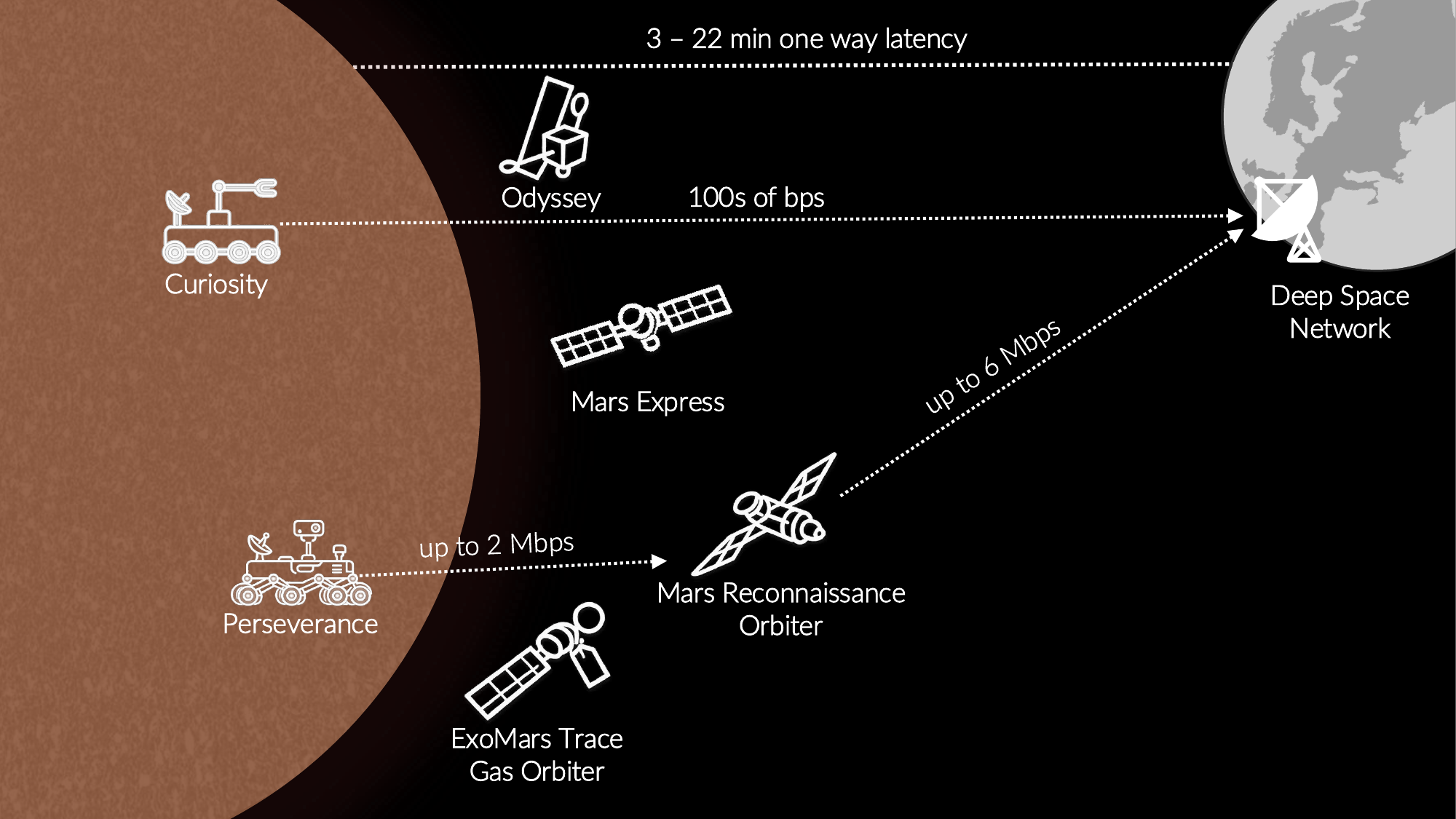}
    \caption{\centering Active Mars missions and their communication with Earth. Surface rovers can communicate directly with Earth at only hundreds of bits per second, or through relay orbiters at up to 2 Mbps that forward data to Earth at up to 6 Mbps. 
    }
    \label{fig:marsfleet}
\end{figure}
Mars currently hosts two active surface rovers: NASA's \textit{Curiosity} in Gale Crater and \textit{Perseverance} in Jezero Crater \cite{nasa_mars_relay_network,nasa_perseverance_mission}. Their high-volume communications are supported by the \textit{Mars Relay Network (MRN)}, which currently comprises \textit{Mars Odyssey}, \textit{Mars Express}, \textit{Mars Reconnaissance Orbiter (MRO)}, and the \textit{ExoMars Trace Gas Orbiter (TGO)}
\cite{nasa_mars_relay_network}. Fig.~\ref{fig:marsfleet} summarizes these missions and their communication with Earth.

The MRN is a scheduled store-and-forward network built from science orbiters, not dedicated communication satellites. During brief UHF contacts, rovers upload buffered data to a passing orbiter, which later forwards it to Earth~\cite{nasa_mrn_guide,nasa_mars_relay_network}. \textit{Perseverance}, for example, can transmit to a nearby orbiter at up to 2\,Mbps, while its direct-to-Earth X-band link supports only hundreds of bits per second~\cite{nasa_perseverance_communications}. MRO then downlinks to Earth at roughly 0.5--4\,Mbps, reaching 6\,Mbps under favorable geometry~\cite{taylor2014mro}. Relay access remains intermittent---roughly a dozen contacts per week---and the MRN provides neither continuous visibility nor programmable compute~\cite{nasa_mars_relay_network}.

NASA-operated relay data reach Earth through the \textit{Deep Space Network (DSN)}, whose antenna complexes in California, Spain, and Australia are shared across interplanetary missions~\cite{nasa_dsn,nasa_mars_relay_network}. A typical science-data path is
$
\text{rover}
\rightarrow
\text{MRN orbiter}
\rightarrow
\text{DSN}
\rightarrow
\text{mission operations center}.
$

The architecture is robust, but intermittent and schedule-driven.

\subsection{Existing Challenges}

\para{Mars instruments generate more data than the Earth link can return.}
Surface assets can transmit to nearby relay orbiters at megabit-per-second rates, but the Mars--Earth backhaul is intermittent and constrained~\cite{nasa_mars_relay_network}. Consequently, missions must prioritize which observations to acquire and return, while potentially useful raw data remain buffered, compressed, or discarded. This mismatch will grow as Mars hosts more capable instruments, vehicles, and sensors. Future infrastructure must therefore increase not only communication capacity, but also the amount of scientific value extracted from each transmitted bit.

\para{Mars--Earth communication limits interactivity and discovery.}
Earth and Mars can be separated by 55-400 million kilometers, imposing one-way propagation delays of roughly 3--22 minutes~\cite{nasa_adaptive_systems}. The operational delay is often longer because of blockages from the Sun or the fact that observations must be returned to Earth, reviewed, and incorporated into a later command plan. Surface mobility reflects this slow control loop: \textit{Curiosity} has traveled about 23 miles since 2012, while \textit{Perseverance} took more than five years to complete its first 26.2 miles~\cite{nasa_curiosity_location,nasa_perseverance_marathon}. These rates also reflect difficult terrain and careful science operations, but most new observations cannot influence rover actions until the next planning cycle.

\para{Mars infrastructure is expensive and difficult to deploy.}
Mars missions cost billions of dollars: \textit{Curiosity}'s estimated life-cycle cost was about \$2.5 billion, while \textit{Perseverance} cost roughly \$2.4 billion to build and launch, plus \$300 million for landing and prime-mission operations~\cite{nasa_oig_msl,nasa_perseverance_presskit}. 
Because Earth--Mars launch opportunities recur only every 26 months, infrastructure cannot be added quickly~\cite{nasa_mars_timeline}. Any orbital-computing architecture must therefore complement existing assets, provide value from its first deployment, and expand across successive launch windows.

\subsection{Related Work}

Terrestrial LEO systems have explored satellite edge platforms and serverless runtimes~\cite{pfandzelter2021leoedge,pfandzelter2024komet}, state placement~\cite{li2022spacecore}, contact-aware scheduling~\cite{tao2023umbra}, split execution for Earth observation~\cite{tao2024serval}, and realistic satellite-network testbeds~\cite{pfandzelter2022celestial,lai2023starrynet}. 
These mechanisms exploit predictable mobility, but their assumptions do not transfer directly to Mars, where infrastructure is sparse, Earth contact is delayed and intermittent, and deployment opportunities are years apart (Table~\ref{tab:leo-vs-mars}).

Mars-specific studies have proposed areostationary spacecraft for persistent communication and monitoring~\cite{montabone2020areostationary} and an 81-satellite low-Mars-orbit edge network for planet-wide coverage~\cite{pfandzelter2023orbitaledge}. Our design starts from a different requirement: shared compute needs a reachable service endpoint, not continuous high-rate access everywhere. We therefore separate persistent reachability from surface link capacity, using a small areostationary tier for shared service and adding low-orbit capacity as demand grows. Our compute layer can be shared across current missions, and future sensors, vehicles and human operations, and can be incrementally deployable providing compute to active Mars missions with just one areostationary node.
\section{A Two-Tier Compute Layer for Mars}

Compute placed on the surface of Mars is tied to a single mission and shares its power, mass, thermal, and lifetime constraints. It also cannot be shared easily across widely separated assets: \textit{Curiosity} and \textit{Perseverance}, for example, operate 3,700 km apart. Mars' orbit provides a natural shared layer because surface data already pass through relay spacecraft on their way to Earth. A compute-bearing orbiter can therefore serve multiple missions, remain useful beyond any single rover or lander, and provide future missions with a standing pool of resources. Orbit also provides a more stable long-term platform than the surface, where dust accumulation has ended missions such as Opportunity and InSight \cite{nasa_opportunity_end, nasa_insight_retired}.

The challenge is that no single Mars orbit provides both persistent access and a strong surface link. We therefore separate these functions across two tiers: areostationary nodes provide persistent shared services, while low-Mars-orbit nodes provide short, high-rate contacts and additional compute.

\subsection{Two Orbital Regimes}

\begin{table}[t]
\centering
\footnotesize
\setlength{\tabcolsep}{3pt}
\renewcommand{\arraystretch}{0.95}
\begin{tabularx}{\columnwidth}{@{}lYY@{}}
\toprule
 & \textbf{Areostationary} & \textbf{Low Mars Orbit} \\
\midrule
Altitude / period
    & $17{,}030$ km / $24.62$ h
    & $400$ km / $118$ min \\
Maximum Eclipse
    & $78.5$ min
    & $41.6$ min per orbit \\
Coverage
    & $33\%$; persistent
    & $2.5\%$; brief passes \\
Minimum Link Distance
    & $17{,}030$ km
    & $400$ km \\
Received Power
    & $1\times$
    & $\sim1{,}800\times$ \\
Scaling
    & geography: 3 reach $\sim$90\%
    & as demand grows \\
Primary Role
    & coverage and shared state
    & latency aware applications and added capacity \\
\bottomrule
\end{tabularx}
\caption{Altitude separates surface coverage from surface-link capacity.}
\label{tab:areo-vs-lmo}
\end{table}

Table~\ref{tab:areo-vs-lmo} summarizes the trade-off between the altitude of nodes in the Mars orbits. An areostationary node remains fixed above the equator at approximately 17{,}030\,km, while a 400\,km low-Mars-orbit node moves rapidly over a much smaller footprint. High altitude increases persistent reachability; low altitude shortens the link to the surface.

With a minimum elevation angle of $\varepsilon_{\min}=10^\circ$ (as used in ~\cite{taylor2014mro}), an areostationary node reaches approximately 33\% of the Martian surface, compared with 2.5\% for a 400\,km node. The low node, however, is about 43$\times$ closer at closest approach. Under identical free-space link assumptions, this corresponds to roughly 1{,}800$\times$ greater received power. Altitude therefore separates two resources: persistent reachability and how much data the surface-link transmits.

\subsection{Three High Nodes Reach Most of Mars}

\begin{figure}[t]
    \centering
    \includegraphics[width=\linewidth]{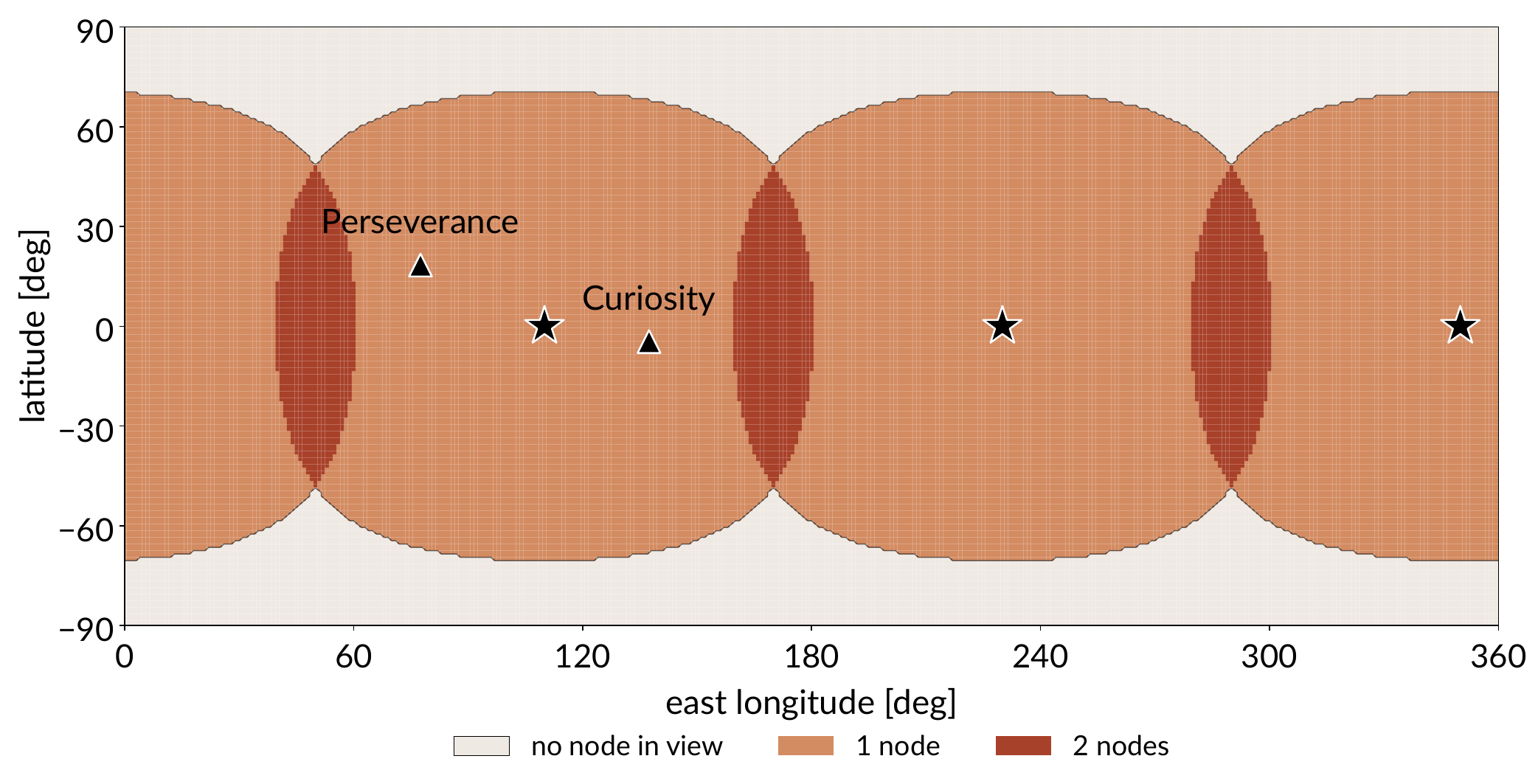}
    \caption{\centering Compute-availability coverage on Mars using areostationary nodes. Each node ($\star$) sees $\pm$70° of Mars. One node covers 33\% of the planet and both active rovers; three nodes at 120° spacing cover 90\% of the surface, leaving only the poles.}
    \label{fig:coverage-map}
\end{figure}

We show the surface coverage of areostationary nodes in Fig.~\ref{fig:coverage-map}. A single areostationary node reaches approximately 33\% of Mars and all the currently active surface rovers on Mars. Three areostationary nodes spaced $120^\circ$ apart extend coverage to roughly 90\% of the planet, with the remaining gaps concentrated near the poles.

This result established the practicality of deploying compute on Mars. A constellation designed for continuous global communication may require tens of satellites. Shared compute has a weaker requirement: a surface asset needs one reachable compute endpoint, not uniform connectivity between arbitrary locations. To do so, only 1 areostationary node is needed to provide compute to all active Mars rovers, and only 2 additional areostationary compute nodes are needed to cover 90\% of Mars. Subsequent addition of high-altitude nodes produce rapidly diminishing coverage returns after the third node. Low-orbit nodes add even less instantaneous coverage because each sees only a small moving region (about 2.5\% of Mars at a time).

Coverage, however, is only half of the problem. The same altitude that gives an areostationary node a broad, persistent footprint also creates a weak surface link. High orbit can make compute reachable, but it cannot efficiently ingest high-volume data from the surface.

\begin{figure}
    \centering
    \includegraphics[width=\linewidth]{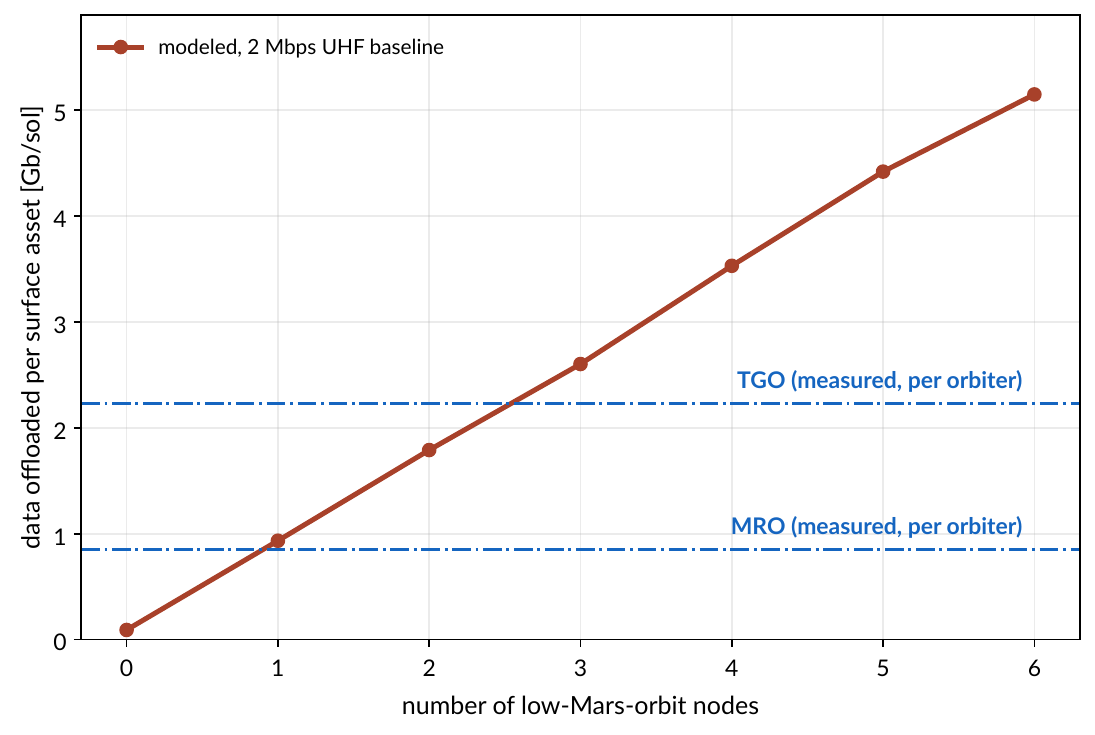}
    \caption{\centering Data offloaded per surface asset versus the number of low-Mars-orbit nodes modelled using 
    the $2$\,Mbps rover-to-orbiter UHF link~\cite{nasa_perseverance_communications}.
    Measured 2026 relay volumes~\cite{nasa_mars_relay_network} are shown for comparison. }
    \label{fig:bandwidth}
\end{figure}

\subsection{Low Orbit Buys Link Quality}

The broad coverage of areostationary nodes makes shared compute reachable, but not necessarily useful for data-intensive applications. A 400\,km node is roughly 40$\times$ closer to the surface than an areostationary node, corresponding to an approximately 1{,}800$\times$ received-power advantage under identical free-space link assumptions. This geometric advantage motivates a low-orbit tier.

We model daily surface-to-orbit offload by evaluating the closest node visible above a $10^\circ$ elevation angle (as used in ~\cite{taylor2014mro}) every minute and assigning rate $r$ as:
\[
r = R_0\left(\frac{D_0}{d}\right)^2,
\]
where $d$ is the \emph{instantaneous slant range} from the rover to that node, $(R_0,D_0)=(2\,\mathrm{Mbps},400\,\mathrm{km})$ is calibrated from Perseverance's published maximum UHF rate~\cite{nasa_perseverance_communications}. Each 400\,km node provides a rover with approximately 2.8 passes per sol, or about 26 minutes of aggregate contact.

Figure~\ref{fig:bandwidth} shows that a rover connected only to an areostationary node (0 low-Mars-orbit nodes) can offload about 0.09\,Gb/sol. Each low-orbit node instead contributes approximately 0.85\,Gb/sol from a few brief passes---roughly ten times the areostationary node's daily surface-ingest capacity. Additional low-Mars-orbit nodes increase offload almost linearly because they contribute largely independent contact opportunities.

Our single-node estimate falls between the relay volumes reported for MRO (0.85\,Gb/day) and TGO (2.23\,Gb/day)~\cite{nasa_mars_relay_network}. The comparison is approximate: NASA reports end-to-end data returned to Earth per orbiter, whereas our model estimates per-rover surface-to-orbit offload under ideal contact utilization. The result should therefore be interpreted as an upper bound rather than an operational prediction.

Unlike high-altitude coverage, low-orbit capacity has no natural saturation point. The required number of low-orbit nodes is determined by how much data the surface fleet generates: additional nodes provide additional contact time and data transfer rates. High-orbit nodes are therefore sized by coverage, whereas low-orbit nodes are sized by demand.

Low orbit therefore complements, rather than replaces, the high-orbit tier by providing a higher bandwidth link that persistent coverage alone cannot.

\subsection{Why Both Tiers Are Needed}

The trade-off becomes clear by comparing three deployments: areostationary-only, low-orbit-only, and two-tier.

\para{Areostationary-only.}
One areostationary node provides persistent access to all active Mars rovers, three nodes cover roughly 90\% of Mars, but the long surface link carries only about 0.09\,Gb/sol under our UHF baseline. This design supports shared state and coordination, but not bulk surface-data transfer.

\para{Low-orbit-only.}
A 400\,km node adds approximately 0.85 Gb/sol of surface ingress, but sees only about 2.5\% of Mars at a time. Near-continuous access would therefore require a much larger constellation, such as the 81-node design explored in prior work~\cite{pfandzelter2023orbitaledge}, and several Earth--Mars launch windows to deploy.

\para{Two-tier.}
A small areostationary tier establishes persistent shared service, while low-orbit nodes are added where greater link bandwidth and near-surface processing are needed. The architecture can be deployed incrementally, with each additional node expanding either coverage or capacity. The two tiers therefore solve different bottlenecks: areostationary nodes establish persistent service, while low-orbit nodes scale surface-data transfer.

\subsection{What Runs Where?}
\label{sec:workloads}

The two tiers support different classes of computation. Low-orbit nodes are closest to surface assets and provide brief, high-rate contacts. They are therefore best suited to data-intensive, latency-sensitive processing: filtering instrument streams, compressing or summarizing observations, detecting events, and selecting which data should be forwarded. Processing data during or shortly after contact reduces the volume that must traverse the Earth-facing links.

Areostationary nodes provide persistent reachability over large regions and are better suited to long-lived shared state. They can maintain maps, scientific models, mission context, and cached datasets that must remain available across rover passes and orbital contacts. They can also combine observations from multiple assets, coordinate follow-up measurements, and expose common services to missions that were not designed to communicate directly.

The two tiers can also form a processing pipeline. A low-orbit node may extract features or identify an unusual observation, while an areostationary node compares it against prior measurements, updates shared state, and determines whether to request additional data or prioritize transmission to Earth. Earth remains responsible for objectives requiring human judgment, global mission planning, and archival science. Mars-local compute instead handles the work whose value depends on acting before the next Earth planning cycle.
\section{Are Martian Compute Nodes Feasible?}
\label{sec:feasible}

We do not claim a flight-ready spacecraft design but instead ask: \emph{does a Mars-orbit compute node pass a first-order feasibility test?} 

We assume a $5$\,kW compute payload on each node and a $15\%$ allowance for avionics, conversion losses, and other spacecraft systems, giving a continuous load of $5.75$\,kW. On Earth, power and cooling are supplied by datacenters. At Mars, they must be carried with the computer, making solar arrays, batteries, radiators, and the spacecraft bus the dominant costs.

\para{Power and thermal control.}
We size solar arrays at a conservative output of $100\,\mathrm{W/m^2}$ (compared to Mars' average solar irradiance of approximately $590\,\mathrm{W/m^2}$) and assume charge and discharge efficiencies of $0.85$ and $0.95$, $70\%$ usable battery depth, a $10$--$15\%$ array margin, and a $10\%$ battery reserve \cite{nasa_mro_presskit,nasa_maven_facts}. A $400$\,km node, with a $118$-minute orbit and a maximum $41.6$-minute eclipse, requires approximately $106$--$110\,\mathrm{m^2}$ of array and a $6.6$\,kWh battery. An areostationary node has a longer maximum eclipse of $78.5$ minutes, but much more sunlight between eclipses, requiring approximately $68$--$71\,\mathrm{m^2}$ of array and a $12.5$\,kWh battery. Low orbit therefore requires more generation capacity, while areostationary orbit requires more energy storage.

Nearly all consumed power becomes heat. Rejecting a $6$\,kW waste-heat load at $300$\,K requires an ideal radiator area of $15.4\,\mathrm{m^2}$. Allowing for imperfect heat transport, external loading, degradation, deployment geometry, and reserve increases the preliminary estimate to $21$--$26\,\mathrm{m^2}$.

\para{Node Mass.}
The spacecraft mass is driven primarily by the systems required for continuous operation. Using NASA-reported technology ranges for power generation and storage, a radiator and heat-transport areal mass of approximately $5$--$10\,\mathrm{kg/m^2}$, and concept-stage assumptions for the compute subsystem and spacecraft bus, we estimate that the solar arrays contribute roughly $140$--$385$\,kg, the radiator and heat-transport system $105$--$260$\,kg, and the battery $45$--$125$\,kg, depending on orbit~\cite{nasa_power_soa_2026,nasa_radiator_mass}. Adding $100$--$160$\,kg for compute and power electronics, $250$--$450$\,kg for communications, propulsion, attitude control, structure, and avionics, and a $20\%$ reserve gives an operational dry mass of approximately $0.8$--$1.6$\,t per node. This places the design in the broad mass class of existing Mars orbiters such as MRO~\cite{nasa_maven_facts,nasa_mro_presskit}.

\para{Incremental deployment.}
Our compute architecture need not be launched as a complete constellation. A first areostationary node provides persistent service to all existing Mars rovers and can be used to validate the compute, thermal, power, and station-keeping systems. Later Earth--Mars launch opportunities, approximately $26$ months apart, can add high-orbit nodes to expand coverage and low-orbit nodes where greater ingest capacity is needed.

As a timely cost reference, SpaceX currently advertises future Starship cargo delivery to the Martian surface at roughly \$100 million per tonne~\cite{spacex_mars_2026}. This is not a quoted price for Mars-orbit insertion, but it places the transport cost of a $0.8$--$1.6$\,t node in the rough \$80--\$160 million range before capture, integration, and mission-specific costs. Each deployment should therefore provide useful capability on its own rather than depend on completing the full constellation.
\section{Open Research Questions for Mars-Local Compute}

We outline some open research questions below.

\para{How does improving onboard AI change Mars systems?} The value of Mars-local compute depends not only on communication constraints but also on what local computation can accomplish. As AI systems become more capable, increasing amounts of planning, scientific interpretation, and anomaly detection may execute near Mars, with Earth consulted only when model confidence is low or human judgment is required. This changes both communication and systems design: bandwidth becomes a resource for exceptions rather than every decision, while runtimes must support confidence-aware execution, model evolution, and verifiable human override across long communication delays. Mars-local compute thus becomes increasingly valuable as local intelligence improves.

\para{Where should computation run?}
Mars offers several execution sites---surface assets, low-Mars-orbit satellites, areostationary nodes, and Earth---with different latency, energy, and connectivity. Systems such as Serval \cite{tao2024serval} and Komet \cite{pfandzelter2024komet} show how predictable mobility can guide split execution and service placement in LEO satellites on Earth. At Mars, however, moving state between nodes may consume the same scarce capacity that computation is meant to save. A Mars runtime must decide when to move code, data, or intermediate results, and when computation should remain ``local''.

\para{How should communication and computation be scheduled together?}
LEO schedulers such as Umbra use predicted contacts to decide when data should be transmitted~\cite{tao2023umbra}. Near Mars, the system must also decide whether to process data before sending it. For example, an orbiter may spend compute and energy to compress an image or detect an important event, thereby reducing its size or increasing its priority. Communication and computation should therefore be scheduled together, rather than treating bandwidth, energy, storage, and processor time as separate resources.

\para{When is approximation safe?}
Semantic filtering and selective inference can increase the useful information returned per downlink bit, as demonstrated by LEO Earth-observation systems such as Serval \cite{tao2024serval}. Mars makes mistakes harder to reverse because an observation may be unique and impossible to reacquire. The default should therefore be to rank, summarize, or progressively encode data rather than silently discard it. The open problem is to define approximation contracts that specify confidence, provenance, retention, and the authority to delete raw observations.

\para{What state should persist near Mars?}
Mars-local compute could maintain shared maps, models, caches, and mission
context across assets that otherwise operate independently. StarCDN shows
that caching reusable content in orbit can reduce reliance on Earth links~\cite{starcdn}. Mars extends
this question beyond replaceable content: maps, scientific products, model
versions, and raw observations may remain authoritative for months or years
while Earth is intermittently unreachable. LEO systems such as SpaceCore
and Komet separate mobile functions from durable state~\cite{li2022spacecore,pfandzelter2024komet},
but Mars cannot assume a continuously reachable source of truth. Raw
observations may need immutable custody records, shared products may require
versioning across long partitions, and transient nodes should only hold 
reconstructible state. The challenge is to provide useful sharing without
losing provenance or creating unsafe consistency assumptions.

\para{Can Mars-local compute use modern accelerators?} In low Earth orbits, commercial hardware can be operated with software hardening (e.g., replication). This is because Earth's magnetosphere shields hardware from radiation. Operation on Mars requires radiation hardening (no magnetosphere). How do the compute-communication tradeoffs change given the comparatively limited capabilities of radiation-hardened hardware?

\para{How should Mars compute fail?}
Dense LEO constellations can often route around failed nodes; sparse Mars infrastructure may lose an entire region or service when one spacecraft fails. Mars-local compute must therefore degrade gracefully into relay-only operation, preserve raw data through missed contacts, and isolate failed instruments or software components without disabling communication.

\section{Conclusion}

Mars exploration is constrained not only by what its instruments can observe, but by how quickly those observations can be returned, processed, and acted upon. We propose a two-tiered compute architecture in the Mars orbit that can fundamentally change how missions operate. A small areostationary tier can establish persistent shared service across most of Mars, while low-orbit nodes can be added where missions need greater data ingress and near-surface processing. The system is useful from its first node, scales incrementally across launch windows, and appears feasible within the mass, power, and thermal envelope of large Mars spacecraft.
The broader opportunity is to treat Mars orbit not only as a relay layer, but as shared computing infrastructure. Doing so could shorten scientific feedback loops, coordinate assets across missions, and make each transmitted bit more valuable before it begins the 400-million-kilometer journey to Earth.

\bibliographystyle{ACM-Reference-Format} 
\bibliography{hotnets25-template}

\end{document}